\documentclass{jnmp}
\usepackage{amsmath}
\JNMPnumberwithin{equation}{section}
\newtheorem{assertion}{Assertion}

\begin{document}

\renewcommand{\evenhead}{A~N~Leznov, J~Escobedo-Alatorre and R~Torres-Cordoba}
\renewcommand{\oddhead}{Fine Structure of the Discrete Transformation}

\thispagestyle{empty}

\FirstPageHead{10}{2}{2003}{\pageref{leznov-firstpage}--\pageref{leznov-lastpage}}{Article}

\copyrightnote{2003}{A~N~Leznov, J~Escobedo-Alatorre and R~Torres-Cordoba}

\Name{Fine Structure of the Discrete Transformation\\
for Multicomponent Integrable Systems}
\label{leznov-firstpage}

\Author{A~N~LEZNOV~$^{\dag^1\dag^2\dag^3}$, J~ESCOBEDO-ALATORRE~$^{\dag^1}$
 and R~TORRES-CORDOBA~$^{\dag^1}$}

\Address{$^{\dag^1}$~Universidad Autonoma del Estado de Morelos, CCICAp,
Cuernavaca, Mexico \\[10pt]
$^{\dag^2}$~Institute for High Energy Physics, 142280, Protvino, Moscow Region, Russia \\[10pt]
$^{\dag^3}$~Bogoliubov Laboratory of Theor. Phys., JINR, 141980 Dubna,
Moscow Region, Russia}

\Date{Received July 24, 2002;
Accepted September 2, 2002}

\begin{abstract}
\noindent
It is shown that in the case of multicomponent integrable systems connected
with algebras $A_n$, the discrete transformation $T$ possesses the fine
structure and can be represented in the form $T=\prod T_i^{l_i}$, where $T_i$
are $n$ commuting basis discrete transformations and $l_i$ are arbitrary
natural numbers. All the calculations are conducted in detail for the case
of a 3-wave interacting system.
\end{abstract}

\section{Introduction}

In the present paper we consider a 3-wave interacting system in the
framework of the discrete transformation theory
\begin{equation}
B_1=-DE^*,\qquad E_2=-DB^*,\qquad D_3=-BE,  \label{MSC}
\end{equation}
where $(D,B,E)$ are 3 complex-valued unknown functions; and $(x_1\equiv 1$,
$x_2\equiv 2$, $x_3\equiv 3)$, three independent arguments of the problem (we
assume them to be real). In the case of $(1+1)$ space, the operators of
differentiation are connected by the additional condition $(\partial_1+
\partial_2+\partial_3=0)$.

The Lax pair for~(\ref{MSC}), in the traditional consideration (by the
inverse scattering method), is connected with $A_2$ algebra~\cite{2}. This
circumstance distinguishes this case from numerous two-component integrable
systems connected with $A_1$ algebra~\cite{4}. Thus, in this example, it is
suitable to investigate the structure of the discrete transformation theory
for multicomponent integrable systems.

The central idea of the method of discrete substitution consists in that,
instead of (\ref{MSC}), a larger system of equations is considered for 6
unknown functions $(D,B,E,Q,P,A)$
\begin{gather}
P_1=-QE,\qquad A_2=-BQ,\qquad Q_3=-PA ,\nonumber\\
B_1=-AD,\qquad E_2=-DP,\qquad D_3=-EB.  \label{MS}
\end{gather}
It is not difficult to see that the additional conditions (of real
character)
\begin{equation}
P=B^*,\qquad A=E^*,\qquad Q=D^*  \label{REA}
\end{equation}
are self-consistent with (\ref{REA}) and are reduced to (\ref{MSC}).

The goal of the present paper is to show that the discrete transformation
of~(\ref{MSC}) can be represented in the form
\begin{equation}
T=T_1^{n_1}T_2^{n_2},  \label{RT}
\end{equation}
where $T_{1,2}$ are two commutative basis transformations, and $n_{1,2}$ are
arbitrary natural numbers. For comparison, we mention that the discrete
transformation in the case of two-component integrable systems has the form
\[
T=T_1^{n}
\]
with the only basis discrete transformation.

To the best of our knowledge, the structure of the discrete transformation
for multicomponent integrable systems was not yet analyzed.

\section{Discrete transformation}

The present section presents the following assertion concerning the symmetry
properties of the system (\ref{MS}).

\begin{assertion}
The system (\ref{MS}) is invariant with respect to three possible changes of
the unknown functions

$T_3$
\begin{gather*}
\bar Q={\frac{1}{D}},\qquad \bar A=-{\frac{B}{D}},\qquad \bar P={\frac{E}{D}}, \\
\bar B=D\left({\frac{B}{D}}\right)_2,\qquad \bar E=-D\left({\frac{E}{D}}\right)_1,\qquad {\frac{\bar
D}{D}}=DQ-(\ln D)_{1,2} ;
\end{gather*}

$T_1$
\begin{gather*}
\bar P={\frac{1}{B}},\qquad \bar Q={\frac{A}{B}},\qquad \bar E=-{\frac{D}{B}}, \\
\bar D=B\left({\frac{D}{B}}\right)_2,\qquad \bar A=-B\left({\frac{A}{B}}\right)_3,\qquad {\frac{\bar
B}{B}}=BP-(\ln B)_{2,3} ;
\end{gather*}

$T_2$
\begin{gather*}
\bar A={\frac{1}{E}},\qquad \bar B={\frac{D}{E}},\qquad \bar Q=-{\frac{P}{E}}, \\
\bar D=-E\left({\frac{D}{E}}\right)_1,\qquad \bar P=E\left({\frac{P}{E}}\right)_3,\qquad {\frac{\bar
E}{E}}=EA-(\ln E)_{1,3} .
\end{gather*}
\end{assertion}

The validity of this assertion can easily be checked by a direct
substitution of the bar quantities into the corresponding equations of
motion and by using the equations of motion for nonbar functions.

In the form presented above, the substitutions $T_i$ can be considered as
mappings connecting six initial (nonbar) functions with six final (bar)
ones. On the other hand, each substitution can be considered as an infinite-
dimensional chain of equations. For instance, the corresponding chain of
equations in the case of $T_1$ substitution has the form
\begin{gather}
{\frac{B^{n+1}}{B^n}}-{\frac{B^n}{B^{n-1}}}=-\left(\ln B^n\right)_{2,3},\qquad D^{n+1}=
B^n\left({\frac{D^n}{B^n}}\right)_2,\qquad A^{n+1}=-B^n\left({\frac{A^n}{B^n}}\right)_3,\!
\nonumber\\
E^{n+1}=-{\frac{D^n}{B^n}},\qquad Q^{n+1}={\frac{A^n}{B^n}}.   \label{CH}
\end{gather}
The first line contains a lattice-like system connecting 3 unknown functions
$(B,D,A)$ at each point of the lattice. The first chain for $B$ functions is
exactly the well-known two-dimensional Toda lattice.

\section{Some properties of the discrete transformations}

All the discrete transformations constructed above are invertible. This
means that a~nonbar unknown function can be represented in terms of the bar
ones. For instance, $T_3^{-1}$ looks as
\begin{gather*}
D={\frac{1}{\bar Q}},\qquad B=-{\frac{\bar A}{\bar Q}},\qquad E={\frac{\bar P}{
\bar Q}}, \\
P=-\bar Q\left({\frac{\bar P}{\bar Q}}\right)_2,\qquad A=\bar Q\left({\frac{\bar A}{\bar Q}}
\right)_1,\qquad {\frac{Q}{\bar Q}}=\bar D\bar Q-(\ln \bar Q)_{1,2}.
\end{gather*}

It is not difficult to check by direct computation that discrete
transformations $T_i$ are commutative $(T_iT_j=T_jT_i)$ on the solutions of
the system~(\ref{MS}).

Below, we present the corresponding calculations to prove that $
T_1T_2=T_2T_1= T_3$. Indeed, the result of the action of $T_1$ on a certain
solution of the system (\ref{MS}) is the following
\begin{gather*}
P^1={\frac{1}{B}},\qquad Q^1={\frac{A}{B}},\qquad E^1=-{\frac{D}{B}}, \\
D^1=B\left({\frac{D}{B}}\right)_2,\qquad A^1=-B\left({\frac{A}{B}}\right)_3,\qquad {\frac{B^1}{B}}
=BP-(\ln B)_{2,3} .
\end{gather*}
The action of the $T^2$ transformation on this solution leads to
\begin{gather*}
A^{21}={\frac{1}{E^1}}=-{\frac{B}{D}},\qquad B^{21}={\frac{D^1}{E^1}}=D\left({
\frac{B}{D}}\right)_2, \qquad Q^{21}=-{\frac{P^1}{E^1}}={\frac{1}{D}},\\
D^{21}=-E^1\left({\frac{D^1}{E^1}}\right)_1= -{\frac{D}{B}}(B(\ln D)_2-B_2)_1=QD^2-D(\ln D)_{12},\\
P^{21}=E^1\left({\frac{P^1}{E^1}}\right)_3={\frac{E}{D}},\qquad E^{21}=\left(E^1\right)^2A^1-E^1
\left(\ln E^1\right)_{13}=-D\left({\frac{E}{D}}\right)_1.
\end{gather*}
The same calculation carried out in the back direction shows that $W^{1,2}=
W^{2,1}=W^3$ -- the result of application of the $T_3$ transformation to an
initial solution~$W$.

Thus, from each given initial solution $W_0\equiv (A,P,Q,E,B,D)$ of the
system~(\ref{MS}), it is possible to obtain the chain of solutions labeled
by two natural numbers ($l_1$, $l_2$, or $(l_3)$) -- the number of times of
applying the discrete transformations $(T_1,T_2,T_3)$ to it (as it was shown
above, $T_1T_2= T_2T_1=T_3$).

The generated chain of equations in $(D,B,E)$ functions is exactly
two-dimensional Toda lattices. Their general solutions in the case of two
fixed ends are well-known~\cite{LS}. As the reader will see below, this fact
allows one to construct many soliton solutions of the 3-wave problem in the
most straightforward way.

\section{Solution of discrete transformation chains}

In the present section, we derive an explicit form of the result of
application of the general discrete transformation $T_1^{n_1}T_2^{n_2}$ to a
specially chosen initial solution of the system~(\ref{MS}).

In the first subsection, we present some necessary equalities connecting the
determinants of the definite form matrices -- the so-called two Jacobi
equalities. In the second subsection, these equalities are used for solution
of the main problem of the present section.

\subsection{Two Jacobi identities}

We begin with the following obvious equalities for determinants of $n$-th
order
\[
{\rm Det}_n(T_n)\equiv D_n\left(\begin{array}{cc} T_{n-1} & a \\
b & \tau \end{array}\right)= D_{n-1}(T_{n-1})\left(\tau-bT^{-1}_{n-1}a\right)\equiv
D_{n-1}(T_{n-1})\tilde \tau,
\]
where $T_{n-1}$ is $(n-1)\times (n-1)$ matrix, $a$, $b$ are
$(n-1)$-dimensional column (row) vectors, respectively, and $\tau$ is a scalar.

For the same reason, the following formula takes place
\[
D_n\left(\begin{array}{ccc} T_{n-2} & a^1 & a^2 \\
b^1 & \tau_{11} & \tau_{12} \\
b^2 & \tau_{21} & \tau_{22} \end{array}\right)
=D_{n-2}(T_{n-2}) D_2\left(\begin{array}{cc}
\tau_{11}-b^1T^{-1}_{n-2}a^1 & \tau_{12}-b^1T^{-1}_{n-2}a^2 \vspace{1mm}\\
\tau_{21}-b^2T^{-1}_{n-2}a^1 & \tau_{11}-b^2T^{-1}_{n-2}a^2 \end{array}\right),
\]
where $a^i$, $b^i$ are $(n-2)$-dimensional column (row) vectors; and
$\tau_{i,j}$ components of a~$2$-di\-mensional matrix. It is evident how
relations of these types can be continued.

Now using the above results, we transform the following expression
\begin{gather*}
D_n\left(\begin{array}{cc} T_{n-1} & a^1 \\
b^1 & \tau_{11} \end{array}\right)
D_n\left(\begin{array}{cc} T_{n-1} & a^2 \\
b^2 & \tau_{22} \end{array}\right)
- D_n\left(\begin{array}{cc} T_{n-1} & a^2 \\
b^1 & \tau_{12} \end{array}\right)
D_n\left(\begin{array}{cc}T_{n-1} & a^1 \\
b^2 & \tau_{21} \end{array}\right)\\
\qquad {}=  D^2_{n-1}(T_{n-1})D_2\left(\begin{array}{cc}
 \tau_{11}-b^1T^{-1}_{n-1}a^1 & \tau_{12}-b^1T^{-1}_{n-1}a^2 \vspace{1mm}\\
\tau_{21}-b^2T^{-1}_{n-1}a^1 & \tau_{11}-b^2T^{-1}_{n-1}a^2 \end{array}\right)\\
\qquad {}=D_{n-1}D_{n+1}\left(\begin{array}{ccc} T_{n-1} & a^1 & a^2 \\
b^1 & \tau_{11} & \tau_{12} \\
b^2 & \tau_{21} & \tau_{22} \end{array}\right).
\end{gather*}
We will treat the last equality as the first Jacobi identity. By the same
technique, it is not difficult to show that the following equality holds
valid:
\begin{gather*}
D_n\left(\begin{array}{cc} T_{n-1} & a^1 \\
b^1 & \tau \end{array}\right)
D_{n+1}\left(\begin{array}{ccc} T_{n-1} & a^1 & a^2 \\
d^1 & \nu & \mu \\
b^2 & \rho & \tau \end{array}\right)\\
\qquad {}- D_n\left(\begin{array}{cc} T_{n-1} & a^1 \\
b^2 & \rho \end{array}\right)
D_{n+1}\left(\begin{array}{ccc} T_{n-1} & a^1 & a^2 \\
d^1 & \nu & \mu \\
b^1 & \tau & \sigma \end{array}\right)\\
\qquad = D_n\left(\begin{array}{cc} T_{n-1} & a^1 \\
d^1 & \nu \end{array}\right) D_{n+1}\left(\begin{array}{ccc}
 T_{n-1} & a^1 & a^2 \\
b^2 & \rho & \tau \\
b^1 & \tau & \sigma \end{array}\right)
\end{gather*}
that is named the second Jacobi identity. These identities can be
generalized to the case of an arbitrary semisimple group. The reader can
find the corresponding results in~\cite{3}.

\subsection{Direct calculations of the discrete transformation chains}

We take an initial solution in the form
\begin{equation}
Q=A=P=0,\qquad B\equiv B(2),\qquad E\equiv E(1),
\qquad D_3=-B E.  \label{IS}
\end{equation}
Application of each of the inverse transformations $T_i^{-1}$ to this
solution is meaningless because of zeroes in denominators (see Section~2).
The chain of equations under this boundary condition will be called as the
chain with a fixed end from the left (from one side).

The result of application of $l_3$ times $T_3$ transformation to such an
initial solution looks as (to check this fact, only two Jacobi identities of
the previous subsection are necessary):
\begin{gather}
Q^{(l_3}=(-1)^{l_3-1}{\frac{\Delta_{l_3-1}}{\Delta_{l_3}}},\qquad
D^{(l_3}=(-1)^ {l_3}{\frac{\Delta_{l_3+1}}{\Delta_{l_3}}},\qquad \Delta_0=1 ,\nonumber\\
A^{(l_3}=(-1)^{l_3}{\frac{\Delta_{l_3}^B}{\Delta_{l_3}}},\qquad P^{(l_3}=
{\frac{\Delta_{l_3}^E}{\Delta_{l_3}}},\qquad \Delta^B_0=\Delta^E_0=0,\nonumber\\
\label{T_3}
B^{(l_3}={\frac{\Delta_{l_3+1}^B}{\Delta_{l_3}}},\qquad E^{(l_3}=(-1)^{l_3} {
\frac{\Delta_{l_3+1}^E}{\Delta_{l_3}}},\qquad \Delta_{-1}=0.
\end{gather}
where $\Delta_n$ are minors of the nth order of an infinite-dimensional
matrix
\begin{equation}
\Delta=\left(\begin{array}{cccc} D & D_2 & D_{22} & \cdots\\
 D_1 & D_{12} & D_{122} & \cdots\\
 D_{11} & D_{112} & D_{1122} & \cdots\\
\cdots & \cdots & \cdots & \cdots\end{array}\right)  \label{DM}
\end{equation}
and $\Delta_{l_3}^E$, $\Delta_{l_3}^B$ are the minors of $l_3$ order in the
matrices of which the last column (or row) is changed to the derivatives of
the corresponding order with respect to argument $1$ of the~$E$ function
(with respect to argument $2$ of the $B$ function in the second case).

In what follows, we will use the notation: $W^{l_3,l_1}$ and ($W^{l_3,l_2}$)
are the result of application of the discrete transformation $T^{l_3}T^{l_1}$
($T^{l_3}T^{l_2}$) to the corresponding component of the 3-wa\-ve field;
$\Delta^{l_3,l_1}$ ($\Delta^{l_3,l_2}$) is the determinant of $l_3+l_1$
($l_3+l_2$) orders, with the following structure of its determinant matrix:
The first $l_3$ rows (columns) of it coincide with the matrix of~(\ref{DM})
and the last $l_1$, $(l_2)$ rows (columns) constructed from the derivatives
of $B$, ($E$) functions with respect to arguments 2 (1).

The result of further application of $l_1$ times $T_1$ transformation to the
solution (\ref{T_3}) looks as
\begin{gather}
P^{(l_3,l_1}={\frac{\Delta_{l_3,l_1-1}}{\Delta_{l_3,l_1}}}, \qquad B^{(l_3,l_1
}={\frac{\Delta_{l_3,l_1+1}}{\Delta_{l_3,l_1}}}, \qquad \Delta_0=1, \qquad
\Delta^{l_3,-1}\equiv \Delta^E_{l_3},\nonumber\\
Q^{(l_3,l_1}=(-1)^{l_3+l_1-1}{\frac{\Delta_{l_3-1,l_1}}{\Delta_{l_3,l_1}}},
\qquad D^{(l_3,l_1}=(-1)^{l_3+l_1}{\frac{\Delta_{l_3+1,l_1}}{\Delta_{l_3,l_1}}
},  \nonumber\\
\label{TT}
E^{(l_3,l_1}=(-1)^{l_3+l_1}{\frac{\Delta_{l_3+1,l_1-1}}{\Delta_{l_3,l_1}}},
\qquad A^{l_3,l_1}=(-1)^{l_3+l_1}{\frac{\Delta_{l_3-1,l_1+1}}{\Delta_{l_3,l_1}}},
\end{gather}
We do not present the explicit form for components $W^{(l_3,l_2}$ that can
be obtained without any difficulties from~(\ref{TT}) by the corresponding
change of the arguments and initial functions~$B$ and~$E$.

\section{Multisoliton solution of the scalar 3-wave problem}

As is mentioned in the Introduction, the system (\ref{MS}) allows the
following reduction (under a further assumption that all operators of
differentiation are the real ones $\partial _\alpha =\partial _\alpha ^{*}$),
given above by (\ref{REA}) ($P=B^{*},\quad A=E^{*},\quad Q=D^{*}$). And
in this case, the system (\ref{MS}) is reduced to the three equations;
$B_1=-DE^{*}$, $E_2=-DB^{*}$, $D_3=-BE$, given above by (\ref{MSC}),
for three complex-valued unknown functions $(E,B,D)$.

Now we would like to demonstrate how the multisoliton solutions of the
system~(\ref{MSC}) can be obtained with the help of the technique of
discrete transformation in the most straightforward way.

With this aim, we consider the action of the direct and inverse
$T_i$, $T_i^{-1}$ transformations on the reduced solution of the system~(\ref{MSC}).
The trick consists in that the discrete transformation does not
preserve the condition of realness (\ref{REA}), and starting with the
solution of the reduced system, we come back to the solution of the
nonreduced one, and only in some special cases, visa versa. We denote the
three-dimensional vector $(Q,P,A)$ by the single symbol $\vec{Q}$ and the
three-dimensional vector $(D,B,E)$ by the symbol $\vec{D}$. Then the result
of action of the direct and inverse transformations on the solution
satisfying the condition of realness $\vec{Q}=\vec{D^*}$ is the following:
\[
T^n_i (\vec{D},\vec{D^*})=(t_i)^n(\vec{q},\vec{d}),\qquad T^{-n}_i (\vec{D},
\vec{D^*})=(\vec{d^*},\vec{q^*})
\]
where $t_i$ are point-like symmetries of the system (\ref{MS})
\begin{gather*}
t_3(Q,P,A,D,B,E)=(Q,-P,-A,D,-B,-E), \\
t_2(Q,P,A,D,B,E)=(-Q,-P,A,-D,-B,E),\\
t_3(Q,P,A,D,B,E)=(-Q,P,-A,-D,B,-E) .
\end{gather*}
It is obvious that $t_i^2=1$. Thus, if we apply the discrete transformation
$2n$ times to the initial bad (nonreduced) solution $(0,\vec{D})$ and, as a
result, obtain $(t^n\vec{D^*} ,0)$, then in the middle of the chain, we will
have a solution satisfying the condition of realness that coincides with an
$n$-soliton solution of the reduced system~(\ref{MSC}).

The solution of the chain with the boundary conditions $\vec{Q}=0$ at the
left end of the chain and $\vec{D}=0$ at the right end will be the chain
with the both fixed ends. Really, the condition $\vec{D}=0$ is the system of
equations, from which the initial functions $D$,
$B$, $E$ (see~(\ref{MSC}) can be
determined as the solutions of ordinary differential equations.

\section{Matrix three-wave problem in the space\\ of three dimensions and its
discrete transformation}

In this section, we will consider an integrable system unknown until now in
the three-dimensional space for 6 unknown functions taking values in some
semisimple algebra. The $L-A$ formalism is not applicable to a system of
that sort, but the formalism of discrete transformations works well. We
emphasize that this system is the local one in all three independent
arguments.

In all the above calculations, we have not used (except for the detailed
resolving of discrete transformation chains) the condition that operators of
differentiation are connected by the expression
\[
\partial_1+\partial_2+\partial_3=0 .
\]
(Provided we do not want to derive explicit multisoliton solutions in the
$(1+1)$ space of the real physical 3-wave problem.)

So, we can consider the system (\ref{MS}) where all three operators are
independent of each other and correspond to differentiation with respect to
one of coordinates of the three-dimensional space. The second generalization
consists in the possibility to treat the unknown function in (\ref{MS}) as
the operator-valued one. Of course, in this case, the order of the
multipliers is essential and should exactly coincide with the one fixed in
the system~(\ref{MS}).

In this case, the following assertion takes place:

\begin{assertion}
The system~(\ref{MS}) with operator-valued unknown functions $(Q,A,P,D,E,B)\!$
is invariant under the following three transformations of the unknown
functions

$T_3$
\begin{gather*}
\bar Q=D^{-1},\qquad \bar A=-BD^{-1},\qquad \bar P=D^{-1}E, \\
\bar B=\left(BD^{-1}\right)_2D,\qquad \bar E=-D\left(D^{-1}E\right)_1,\qquad D^{-1}\bar
D=QD-\left(D^{-1}D_2\right)_1 ;
\end{gather*}

$T_1$
\begin{gather*}
\bar P=B^{-1},\qquad \bar Q=B^{-1}A,\qquad \bar E=-D B^{-1}, \\
\bar D=\left(D B^{-1}\right)_2 B,\qquad
\bar A=-B\left(B^{-1}A\right)_3,\qquad \bar B B^{-1}=BP-\left(B_3B^{-1}\right)_2 ;
\end{gather*}

$T_2$
\begin{gather*}
\bar A=E^{-1},\qquad \bar B=E^{-1} D,\qquad \bar Q=-P E^{-1}, \\
\bar D=-E\left(E^{-1}D\right)_1,\qquad \bar P=\left(P E^{-1}\right)_3 E,\qquad E^{-1}\bar
E=AE-\left(E^{-1}E_3\right)_1 .
\end{gather*}
\end{assertion}

The validity of this assertion can be verified by a direct rather simple
calculation, like in Section~2.

As in the scalar case, these discrete transformations of the functions under
consideration are commutative. The arising chains of equations for $(E,B,D)$
operator-valued functions (the matrices of finite dimensions, for instance)
coincide with the matrix Toda chains investigated above. Explicit solutions
to these chains of equations with the fixed ends can be found in~\cite{LY}.
Combining these results, it is possible to construct multisoliton solutions
of the matrix 3-wave problem in three dimensions in a way similar to that
proposed in~\cite{LYY} for construction of multisoliton solutions to the
matrix Devay--Stewartson equation.

\section{Outlook}

The concrete results of the present paper consist in deriving explicit
formulae of discrete transformations for the 3-wave problem (Section~2) and
their generalization to the matrix case (Section~6).

However, it is not less important to understand how the method of the
discrete transformations can be generalized to the case of multicomponent
systems connected with the semisimple algebras of higher ranks~$r$. From
results of the present paper it is clear that in the case of an arbitrary
semisimple algebra, there are $r$ independent basis commuting discrete
transformations. How these commutative objects are connected with the main
ingredients of the representation theory of groups is a very interesting and
intriguing question for further investigation.

And the last comment: The chain with two fixed ends cannot be considered as
the basis for some finite-dimensional representation of the group of the
discrete transformations, if at all, it is possible to use the term ``group''
in this case. On the basis function at the end point of the chain on the
right, it is impossible to act by a direct discrete transformation and by an
inverse transformation at the left end. What is the discrete transformation
from the group-theoretical point of view in this case? At present, we have
no answer to this question.

\subsection*{Acknowledgments}

The authors are indebted to CONACYT for financial support.

\label{leznov-lastpage}


\begin{thebibliography}{9}
\small

\bibitem{2}  Zakharov V~E, Manakov~S~M, Novikov~S~P and
Pitaevskii~L~P, Theory of Solitons. The Method of the Inverse
Scaterring Problem, Moscow, Nauka, 1980 (in Russian).

\bibitem{4}  Leznov~A~N, Integrable systems,
{\it Fiz. Elem. Chast. Atom. Yadra} {\bf 27}, Nr.~5
(1996), 1161--1246.

\bibitem{LS}  Leznov~A~N and Saveliev M~V, Group Theoretical
Methods for Integration of Nonlinear Dynamical Systems, Progress in Physics
15, Birkhauser, Basel, 1992.

\bibitem{3}  Leznov~A~N, A New
Approach to the Theory of Representations of Semisimple Lie Algebras and
Quantum Algebras, {\it Theor. Math. Phys.} {\bf 123} (2000),
633--650.

\bibitem{LY}  Leznov~A~N and Yusbashjan~E~A, The General
Solution of Two-Dimensional Matrix Toda Chain Equations with Fixed
Ends, {\it Lett. Math. Phys.}
{\bf 35} (1995), 345--349.

\bibitem{LYY}  Leznov~A~N and Yusbashjan E~A, Multi-Soliton
Solutions of the Two-Dimensional Matrix Davey--Stewartson Equation, {\it Nucl. Phys.} {\bf B496}
(1997), 643--653.
\end{thebibliography}
\end{document}